**Perspective**

# Silicon is the next frontier in plant synthetic biology

Aniruddha Acharya[a✉], Kaitlin Hopkins[b] & Tatum Simms[c]

[a]Biological and Earth Sciences Department; Arkansas Tech University, Russellville, AR, USA 72801

[b]School of Agricultural Sciences; Sam Houston State University, Huntsville, TX, USA 77341

[c]Department of Agriculture and Tourism; Arkansas Tech University, Russellville, AR, USA 72801

✉ aniruddha1302@gmail.com

Work email – aacharya2@atu.edu

ORCiD - 0000-0002-6774-5934

**Author Contributions Statement**

Conceptualization and writing original draft A.A. Reviewing, editing, formatting K.H., T.S., and A.A. All authors have read and agreed to the submitted version of the manuscript.

**Abstract**

Silicon has striking similarity with carbon and is found in plant cells. However, there is no specific role that has been assigned to silicon in the life cycle of plants. The amount of silicon in plant cells is species specific and can reach levels comparable to macronutrients. Silicon is the central element for artificial intelligence, nanotechnology and digital revolution thus can act as an informational molecule like nucleic acids while the diverse bonding potential of silicon with different chemical species is analogous to carbon and thus can serve as a structural candidate such as proteins. The discovery of large amounts of silicon on Mars and the moon along with the recent developments of enzyme that can incorporate silicon into organic molecules has propelled the theory of creating silicon-based life. More recently, bacterial cytochrome has been modified through directed evolution such that it could cleave silicon-carbon bonds in organo-silicon compounds thus consolidating on the idea of utilizing silicon in biomolecules. In this article the potential of silicon-based life forms has been hypothesized along with the reasoning that autotrophic virus-like particles can be a lucrative candidate to investigate such potential. Such investigations in the field of synthetic biology and astrobiology will have corollary benefit on Earth in the areas of medicine, sustainable agriculture and environmental sustainability. Bibliometric analysis indicates an increasing interest in synthetic biology. Germany leads in research related to plant synthetic biology, while Biotechnology and Biological Sciences Research Council (BBSRC) at UK has highest financial commitments and Chinese Academy of Sciences generates the highest number of publications in the field.

**Keywords**: silicon, synthetic biology, directed evolution, space colonization, autotrophs

Abbreviations:

NIPs: Nod 26-like intrinsic proteins

IPNI: International Plant Nutrition Institute

LTR: Long Terminal Repeat

CRISPR: Clustered Regularly Interspaced Short Palindromic Repeats

NCBI: National Center for Biotechnology Information

**Introduction to Silicon**

Life as we know so far is mainly made of four elements namely, carbon (C), hydrogen (H), oxygen (O) and nitrogen (N), while some other light elements are required in small amounts (White & Brown 2010). It is interesting to note that silicon (Si) in spite of being the second most abundant element on Earth's crust and having striking chemical similarity with carbon does not find itself in the list of elements that is required for life. Silicon is tetravalent and can bond with various chemical species, thus, it is chemically the closest analogue of carbon. This has attracted the imagination of scientists for silicon-based life. The idea has received significant attention with respect to space colonization after the discovery of silicon as being the second most abundant element on the moon and Mars where carbon is in extremely low amounts (Jacob, D. T. 2016). The excitement is not unfounded because though silicon is not deemed as required by lifeforms, the existence of silicon is found in various living organisms including diatoms and plants (Kröger and Poulsen, 2008; Frew et al. 2018).

**Silicon and Plant Physiology**

Plant physiologists do not consider silicon as an essential element for growth and development of plants even though the amount of silicon in several plant species is comparable to many macronutrients. Plants sequester silicon in the form of silicic acid $Si(OH)_4$ and polymerize them to silica that contributes to the rigidity of aboveground plant parts (Exley 2015; Coskun et al. 2019). The deposition of silicon in apoplastic barriers was previously thought to be a passive physiological byproduct. However, recent reports indicate the intricate role of proteins and signaling in such deposits (Mandlik et al. 2020). The symplastic transport of $Si(OH)_4$ involves aquaporins such as Nod 26-like intrinsic proteins or NIPs (Gregoire et al. 2012; Deshmukh et al. 2013; Exley 2015). About 0.1% to 10% of total dry weight of plants can be constituted of silicon, several studies have indicated the phytoprotective role of silicon in plants with respect to countering abiotic and biotic stress (Epstein 1994; Epstein 1999; Adrees etal. 2015; Rizwan etal. 2015). Besides its role in abiotic and biotic stress management, the element is also involved in structural roles in cell wall and is hypothesized to have a signaling role that impacts cellular metabolism, it alters metabolism in plants through modulating endogenous phytohormones (Kim et al. 2014; Luyckx et al. 2017). Silicon imparts mechanical strength to the cell wall of plants and is found in the form of silica ($SiO2$). Many pteridophytes and monocot have noticeable deposits of silicon (Guerriero et al. 2016). Silicon can bind with hydroxyl groups of sugars to form silicate esters and has been classified according to the biochemical and physiological role in plants (Pavlovic et al. 2021). Due to the beneficial roles of silicon in plants, the International Plant Nutrition Institute (IPNI) recently classified it as a beneficial element (Coskun et al. 2019).

**Silicon: From abiogenic to biogenic**

Due to its non-reactive nature, silicon is biocompatible and is widely used in medical devices, pharmaceuticals and in the agricultural industry. The foundation of digital revolution, computation, electronics and artificial intelligence is attributed to silicon. The element is one of the prime candidates for chemical and green synthesis of nanoparticles which can be further utilized in many biological processes. For the abovementioned reasons, it can be stated empirically that silicon can serve in structural, functional and informational roles in some ways as does protein, RNA and DNA respectively. Though the element itself and in a solitary capacity might not be able to achieve such complex role as the biological macromolecules, however, it can bind with other organic and inorganic molecules to achieve such activity. The current literature supports such evidence where silicon is attributed to structural (cell wall), functional (cell signaling) and informational (semiconductor) roles (Guerriero et al. 2016; Wang et al. 2022; Shanmugaiah et al. 2023). Thus, it will not be an exaggeration to hypothesize that silicon can serve as the element that can serve as a key matrix for synthetic life besides being a useful element for sustaining life on Mars and the moon. Silicon bonds being weaker than carbon bonds might add a lot of flexibility to the repertoire of chemical groups that it can bind when the right conditions are provided (Petkowski et al. 2020). Thus, it can be utilized to design synthetic particles that can have virus-like properties and would be an intermediate lifeform. Such particles must have the capability to capture energy, multiply vigorously and convert inorganic carbon to biogenic carbon compounds. Thus, analogous to the oxygenation of Earth, silicon-based life can be utilized for terraforming Mars and the moon to create conditions that would be conducive for the propagation of autotrophs. Viruses can infect autotrophs which indicates that it can utilize the cellular machinery of autotrophs to produce viral particles. Thus, it might be possible to create virus particles that are autotrophic in nature and similar to cyanobacteria can terraform planets to make them Earth-like through synthesis of biogenic carbon. Viruses are obligate intracellular parasites, thus virus-like particles as hypothesized above would need a host for their survival. This challenge can be circumvented if such particles can multiply in a chemical environment such as the "primordial soup" which has adequate chemical diversity resulting in thermodynamically stable reactions leading to the formation of complex molecules. The existence of LTR (long terminal repeat) transposons and their significant implications in plant genome evolution (He et al. 2024) indicates that the idea of virus-autotroph interactions resulting in paradigm shift of life forms is not far-fetched.

The progress in artificial intelligence, metabolic engineering, computation, synthetic biology and genome editing techniques such as CRISPR has propelled the promise of artificial life. Such technologies can be integrated and their potential can be harnessed to modulate life forms and propel the terraforming of extraterrestrial planets. The biology of transposons, prions, viroid, virus, archaea and extremophiles makes it obvious that besides the conventional understanding of life, there are powerful anomalies that reflects the 4 billion years of the history of life on Earth. The scientific community has accepted the theory of abiogenic origin of life. Thus, it is plausible that synthetic life could be made for terraforming by utilizing the powerful scientific tools at our disposal and the longstanding scientific knowledge of the past 300 years. The chemical signatures of water molecules on the moon and Mars along with the current developments in spaceflight technologies offer encouragement that an ancient primordial soup can be created that may be utilized as a matrix for the directed evolution of synthetic life forms. The idea proposed here is different from the conventional hypothesis of silicon-based life because instead of directly creating silicon-based plants and animals, here, the role of an intermediate lifeform is proposed where silicon will play a key role. Such particles could sequester and enrich carbon on the surface of the planets. The sequestration of carbon would establish and accelerate conditions required for carbon-based autotrophic life on the uninhabited planets. The strength of this hypothesis resides in the fact that a more gradual route of directed evolution is proposed that seems more likely to procreate than the challenges inherent to creating silicon-based plants and animals. Besides, virus-like particles would be more amenable to the harsh environment than cellular lifeforms and their vigorous multiplication would accelerate the process of conversion of inorganic carbon to biogenic forms.

**Genetic Engineering and Synthetic Life**

Synthetic genome has been utilized to reconstruct virus, bacteria and eukaryotic cells leading to breakthrough discoveries in medicine, agriculture and fundamental understanding of cells (Venter et al. 2022). Nonstandard amino acids have been incorporated into proteins thus enhancing chemical functionalities of enzymes and expanding the genetic code (Drienovská & Roelfes 2020). Synthetic biology has made significant advances such that the theoretical concepts of the resurrection of extinct plants and animals are actively being actively tested (Shapiro 2017; Albani et al. 2022). Barring rare exceptions, biomolecules found in living organisms maintains a chirality that is represented by D-sugar and L-amino acids. However, mirror-image macromolecules such as nucleic acids and functional proteins have been synthesized with altered chirality (Xu & Zhu 2022). This supersedes the incremental stages of molecular evolution and is a significant step towards creation of mirror cells, however, it has subsequently raised concerns regarding their potential impact on health and environment (Adamala et al. 2024). Though scientists are skeptical, however, without further research, it will be preemptive to conclude the potential and dangers of mirror cells. It is increasingly evident from the early genetic engineering technology such as terminator technology (Niiler 1999) to the most recent CRISPR gene editing technology (Adli 2018) that biological organisms can be manipulated and their propagation can be controlled. Our understanding of telomeres, telomerase, epigenetics and selective toxicity of antibiotics reinforces that genetic engineering and synthetic biology is propelling towards innovations that can be used to regulate life precisely. Thus silicon-based acellular or synthetic microbial life forms may become a reality that can accelerate terraforming besides having applications in agriculture, medicine, environment, fuel, textiles and beyond. Such microbial life can be bioengineered towards self-destruction as a safeguard to prevent pandemic or environmental catastrophe.

Genetic engineering is utilized efficiently to manipulate life (Gibson et al. 2010); however, if we think of creating novel lifeforms then an interdisciplinary approach such as synthetic biology seems more plausible. The fundamental basis of life is inorganic elements thus the macromolecular building blocks of life depends on inorganic elements. Thus, it is important to use a multipronged approach involving biology, chemistry, physics and artificial intelligence to generate synthetic lifeforms where silicon can replace or complement carbon as the predominant element. Since silicon is abundantly available on Mars and the moon, it can be utilized to create molecules for structural, enzymatic and informational roles or can be assimilated with carbon to form organosilicon compounds with such properties. Corollary benefit of such intelligent and synthetic lifeforms might be reflected on Earth where they could be utilized for a multitude of purposes including environmental reclamation, medical breakthroughs and agricultural improvements. Researchers in the California Institute of Technology were able to synthesize organo-silicon compounds by mutating cytochrome c from bacteria. The mutation resulted in incorporation of silicon into hydrocarbons and the results were replicable in vivo under physiological conditions (Kan et al. 2016). Conversely, recent reports include the engineering of an enzyme (bacterial cytochrome P450BM3) through directed evolution that can cleave the silicon-carbon bonds in linear and cyclic volatile methylsiloxanes leading to another major progress in organosilicon chemistry (Sarai et al. 2024). The abovementioned discoveries (Kan et al. 2016 & Sarai et al. 2024) illustrates the potential of enzymes for silicon metabolism in prokaryotes, thus, will stimulate the investigation and manipulation of several enzymes in prokaryotes and eukaryotes concerning utilization of silicon in biomolecules. Such work of directed evolution can pave the way for silicon-based synthetic life. Highly selective and efficient reactions mediated by biocatalyst have been developed with high synthetic value. Such methods can be utilized for synthesis and degradation of organosilicon molecules (Sarai et al. 2021).

**Recent Trends in the use of Silicon in SynBio**

To analyze trends (Figure 1) of scientific publications concerning silicon in relation to synthetic biology, a search was conducted using the "PubMed Advanced Search Builder" application of NCBI PubMed database (NCBI 2024). Two searches were run using the "All Fields category". The first search was a comprehensive search using the search term silicon along with artificial life, artificial cell, synthetic life, synthetic cell and synthetic biology. The query box reflecting the search paradigm was "((((((silicon)) AND (artificial life)) OR (artificial cell)) OR (synthetic life)) OR (synthetic cell)) OR (synthetic biology)". The second search was a conservative search using more restrictive search terms that exclusively reflected the use of silicon in relation to biology. The search terms used for the conservative search were silicon along with artificial life, synthetic life and synthetic biology. The query box reflecting the search paradigm for the conservative search was "(((silicon) AND (artificial life)) OR (synthetic life)) OR (synthetic biology)". Both searches included results of publication numbers from 1980 to 2024 and yielded 255,650 results for the comprehensive search and 97,717 results for the conservative search. Silicon is widely used in biological, medical, chemical and material research; however, our interests are concerning the use of silicon in synthetic biology. Thus, it was important to distinguish the biological and nonbiological use of silicon in the scientific literature. This was necessary to avoid overrepresentation or inflation of the data points pertaining to the use of silicon in biological/medical research. The NCBI PubMed database was selected for the investigation because it is freely available and maintains data mostly pertaining to medical and biological research. Thus, it is advantageous as it eliminates most results that are of nonbiological origin while representing a vast repertoire of biological research conducted across the globe. The search results indicate a sharp increase in the number of publications related to silicon/synthetic biology in the past 25 years. The trendline for both comprehensive and conservative data shows similar pattern with their respective $R^2$ value being 0.862 and 0.724. This reflects that there is an upward trend in research related to silicon and that includes research related to the use of silicon in synthetic biology. Though the perspective presented here is novel, however, on a broader sense it is a shared perspective and the data shows that it is being actively tested across the globe. Furthermore, another search was conducted using the "PubMed Advanced Search Builder" application of NCBI PubMed database (NCBI 2024) in the "All Fields" category and using the search term "Plant Synthetic Biology". The results retrieved from the database representing a timeline from 2010 to 2024 were curated manually based on different categories (Table 1). The results indicate that besides the US government, the non-US government is heavily invested in this field of research.

**Silicon and Plant SynBio**

Silicon alleviates abiotic and biotic stress in plants indicating empirical evidence of its unsuspected role in plant physiology (Frew et al. 2018). Though, initially the protective role of silicon in plants was attributed to its deposition in cell wall, however, later their role in hormonal cross-talk and metabolism was confirmed and the use of silicon nanoparticles to improve commercially valuable traits of crops is explored (Luyckx et al. 2017). Besides plants, biosilicification is reported in several invertebrates and marine organisms, thus, investigating the molecular mechanism of such processes could offer biotechnological candidates for manufacturing of silicon-based commercial products (Morse 1999). Plant synthetic biology is an emerging field with a history of nearly 20 years and holds significant promise in several areas including global food security and clean energy (Joshi & Hanson 2024). It employs engineering of metabolic pathways to elicit user-tailored plant traits. It is considered the next frontier in plant biology where physiological and developmental pathway of plants are reprogrammed for production of adequate food, feed, fuel and pharmaceuticals with minimum inputs thus propelling sustainable environment and agriculture (Fesenko & Edwards 2014; Baltes & Voytas 2015). A document search in Web of Science was performed in March, 10th, 2025. The search category was set as "Topic" and the search term used was "Plant Synthetic Biology". The category "Topic" includes title, abstract and keywords. Search results retrieved 2,603 publications selected from Web of Science Core Collection. Result analysis indicated an increasing interest in the subject over the past 10 years period (Figure 2 a). Germany leads among all countries in research related to plant synthetic biology (Figure 2 b), while Biotechnology and Biological Sciences Research Council (BBSRC) at UK is the organization that leads among the funding agencies to support research in the area (Figure 2 c). Chinese Academy of Sciences generates the highest number of publications in the field (Figure 2 d). The current trend in silicon research and the progress in synthetic biology makes it obvious that the amalgamation of both towards a product or process of human value is imminent. It is imperative that silicon has much larger role in plants that previously expected. Due to several regulatory laws that are essential for animal research, plants are an excellent model to estimate the effect of silicon-based biomolecules in eukaryotic cells. Conversely, the silicon enriched product synthesized within plant cells can be harvested and used to analyze their effect with regards to human consumption and use.

**Conclusion**

The detection of signatures for life in the form of microorganisms or organic compounds has been explored through several technological developments and investigation of topographies in Earth and extraterrestrial entities (Guzman et al. 2020; Enya et al. 2022; Weng et al. 2022). Lava tube speleothems are considered to have analogous environments as compared to Mars and the moon, thus, the molecular, chemical, biochemical and mineralogical investigation of such environments may gather insights of potential extraterrestrial biosignatures (Palma et al. 2024). The potential of silicon to complement carbon or serve as a major element for development of lifeforms that can terraform Mars and other extraterrestrial entities to habitable environments will reinforce such astrobiological efforts of pursuing life beyond Earth. The role of silicon in ameliorating biotic and abiotic stress in plants along with the discovery of silicon transporters in plant cells indicates that the element might have an unsuspected physiological role in plant life (Gaur et al. 2020; Ranjan et al. 2021). Recent reports classify silicon as an essential micronutrient for some plants (Majeed et al. 2012). The demands for microgreen are increasing and with the development of modern agriculture technologies such as precision agriculture, climate-smart agriculture, hydroponics and artificial intelligence augmented agriculture, it is possible to investigate the role of silicon in mass production of microgreens. Estimating the effects of silicon in the mass production of such edible plant products in a controlled environment will help scientists better understand the role of the element in securing food sustainability in contained spaces such as space station and habitats in Mars and the moon. The role of silicon in improving nutritive value of edible plant products needs to be assessed more carefully. Recent discovery of enzymes that can incorporate or cleave silicon in organosilicon compounds holds immense promise in the use of this element in biogenic heterogenous molecules. Plant synthetic biology is an emerging field and the use of silicon in producing mirror cells or the reprogramming of plant cells for the synthesis of silicon-enriched product of commercial value is imminent. The element should receive more attention from plant scientists to investigate the physiological and molecular role of the element in plant life. Such investigations may lead to the development of silicon rich resilient crops on Earth and autotrophic life forms in space thus welcoming silicon as the new green!

**Acknowledgement**

We thank the anonymous reviewers for their insights.

**Statements and Declarations**

**Competing Interests:** The authors have no financial or non-financial interests that are directly or indirectly related to the work submitted for publication.

**Funding:** This work and the publication cost is supported by Department of Biological & Earth Sciences at Arkansas Tech University

**References**

Adamala, K. P., Agashe, D., Belkaid, Y., Bittencourt, D. M. D. C., Cai, Y., Chang, M. W., ... & Zuber, M. T. (2024). Confronting risks of mirror life. *Science*, eads9158.

Adli, M. (2018). The CRISPR tool kit for genome editing and beyond. *Nature communications*, *9*(1), 1911.

Adrees, M., Ali, S., Rizwan, M., Zia-ur-Rehman, M., Ibrahim, M., Abbas, F., ... & Irshad, M. K. (2015). Mechanisms of silicon-mediated alleviation of heavy metal toxicity in plants: a review. *Ecotoxicology and Environmental Safety*, 119, 186-197.

Albani Rocchetti, G., Carta, A., Mondoni, A., Godefroid, S., Davis, C. C., Caneva, G., ... & Abeli, T. (2022). Selecting the best candidates for resurrecting extinct-in-the-wild plants from herbaria. *Nature Plants*, 8(12), 1385-1393.

Baltes, N. J., & Voytas, D. F. (2015). Enabling plant synthetic biology through genome engineering. *Trends in biotechnology*, 33(2), 120-131.

Coskun, D., Deshmukh, R., Sonah, H., Menzies, J. G., Reynolds, O., Ma, J. F., ... & Bélanger, R. R. (2019). In defence of the selective transport and role of silicon in plants. *The New Phytologist*, 223(2), 514-516.

Deshmukh, R. K., Vivancos, J., Guérin, V., Sonah, H., Labbé, C., Belzile, F., & Bélanger, R. R. (2013). Identification and functional characterization of silicon transporters in soybean using comparative genomics of major intrinsic proteins in Arabidopsis and rice. *Plant molecular biology*, 83, 303-315.

Drienovská, I., & Roelfes, G. (2020). Expanding the enzyme universe with genetically encoded unnatural amino acids. *Nature Catalysis*, 3(3), 193-202.

Gaur, S., Kumar, J., Kumar, D., Chauhan, D. K., Prasad, S. M., & Srivastava, P. K. (2020). Fascinating impact of silicon and silicon transporters in plants: A review. *Ecotoxicology and Environmental Safety*, 202, 110885.

Gibson, D. G., Glass, J. I., Lartigue, C., Noskov, V. N., Chuang, R. Y., Algire, M. A., ... & Venter, J. C. (2010). Creation of a bacterial cell controlled by a chemically synthesized genome. *Science*, 329(5987), 52-56.

Grégoire, C., Rémus-Borel, W., Vivancos, J., Labbé, C., Belzile, F., & Bélanger, R. R. (2012). Discovery of a multigene family of aquaporin silicon transporters in the primitive plant Equisetum arvense. *The Plant Journal*, 72(2), 320-330.

Guerriero, G., Hausman, J. F., & Legay, S. (2016). Silicon and the plant extracellular matrix. *Frontiers in plant science*, 7, 463.

Guzman, M., Szopa, C., Freissinet, C., Buch, A., Stalport, F., Kaplan, D., & Raulin, F. (2020). Testing the capabilities of the Mars Organic Molecule Analyser (MOMA) chromatographic columns for the separation of organic compounds on Mars. *Planetary and Space Science*, *186*, 104903.

Enya, K., Yoshimura, Y., Kobayashi, K., & Yamagishi, A. (2022). Extraterrestrial life signature detection microscopy: Search and analysis of cells and organics on Mars and other solar system bodies. *Space Science Reviews*, 218(6), 49.

Epstein, E. (1999). Silicon. *Annual review of plant biology*, 50(1), 641-664.

Exley, C. (2015). A possible mechanism of biological silicification in plants. *Frontiers in Plant Science*, 6, 853.

He, Bing, Wanfei Liu, Jianyang Li, Siwei Xiong, Jing Jia, Qiang Lin, Hailin Liu, and Peng Cui. "Evolution of Plant Genome Size and Composition." *Genomics, Proteomics & Bioinformatics* (2024): qzae078.

Jacob, D. T. (2016). There is no Silicon-based Life in the Solar System. *Silicon*, 8, 175-176.

Fesenko, E., & Edwards, R. (2014). Plant synthetic biology: a new platform for industrial biotechnology. *Journal of Experimental Botany*, 65(8), 1927-1937.

Frew, A., Weston, L. A., Reynolds, O. L., & Gurr, G. M. (2018). The role of silicon in plant biology: a paradigm shift in research approach. *Annals of botany*, 121(7), 1265-1273.

Joshi, J., & Hanson, A. D. (2024). A pilot oral history of plant synthetic biology. *Plant Physiology*, 195(1), 36-47.

Kan, S. J., Lewis, R. D., Chen, K., & Arnold, F. H. (2016). Directed evolution of cytochrome c for carbon–silicon bond formation: Bringing silicon to life. *Science*, 354(6315), 1048-1051.

Kim, Y. H., Khan, A. L., Kim, D. H., Lee, S. Y., Kim, K. M., Waqas, M., ... & Lee, I. J. (2014). Silicon mitigates heavy metal stress by regulating P-type heavy metal ATPases, Oryza sativa low silicon genes, and endogenous phytohormones. *BMC plant biology*, 14, 1-13.

Kröger, N., & Poulsen, N. (2008). Diatoms—from cell wall biogenesis to nanotechnology. *Annual review of genetics*, 42(1), 83-107.

Luyckx, M., Hausman, J. F., Lutts, S., & Guerriero, G. (2017). Silicon and plants: current knowledge and technological perspectives. *Frontiers in plant science*, 8, 411.

Majeed Zargar, S., Ahmad Macha, M., Nazir, M., Kumar Agrawal, G., & Rakwal, R. (2012). Silicon: A Multitalented Micronutrient in OMICS Perspective–An Update. *Current Proteomics*, 9(4), 245-254.

Mandlik, R., Thakral, V., Raturi, G., Shinde, S., Nikolić, M., Tripathi, D. K., ... & Deshmukh, R. (2020). Significance of silicon uptake, transport, and deposition in plants. *Journal of Experimental Botany*, 71(21), 6703-6718.

Morse, D. E. (1999). Silicon biotechnology: harnessing biological silica production to construct new materials. *TRENDS in Biotechnology*, 17(6), 230-232.

National Center for Biotechnology Information (NCBI)[Internet]. Bethesda (MD): National Library of Medicine (US), National Center for Biotechnology Information; [1988] – [cited 2024 Dec 19. Available from: https://www.ncbi.nlm.nih.gov/

Niiler, E. (1999). Terminator technology temporarily terminated. *Nature Biotechnology*, *17*(11), 1054-1054.

Palma, V., De la Rosa, J. M., Onac, B. P., Sauro, F., Martínez-Frías, J., Caldeira, A. T., ... & Miller, A. Z. (2024). Decoding organic compounds in lava tube sulfates to understand potential biomarkers in the Martian subsurface. *Communications Earth & Environment*, 5(1), 530.

Pavlovic, J., Kostic, L., Bosnic, P., Kirkby, E. A., & Nikolic, M. (2021). Interactions of silicon with essential and beneficial elements in plants. *Frontiers in Plant Science*, 12, 697592.

Petkowski, J. J., Bains, W., & Seager, S. (2020). On the potential of silicon as a building block for life. *Life*, 10(6), 84.

*PubMed*. (n.d.). PubMed. https://pubmed.ncbi.nlm.nih.gov/

Ranjan, A., Sinha, R., Bala, M., Pareek, A., Singla-Pareek, S. L., & Singh, A. K. (2021). Silicon-mediated abiotic and biotic stress mitigation in plants: Underlying mechanisms and potential for stress resilient agriculture. *Plant Physiology and Biochemistry*, 163, 15-25.

Rizwan, M., Ali, S., Ibrahim, M., Farid, M., Adrees, M., Bharwana, S. A., ... & Abbas, F. (2015). Mechanisms of silicon-mediated alleviation of drought and salt stress in plants: a review. *Environmental Science and Pollution Research*, 22, 15416-15431.

Sarai, N. S., Levin, B. J., Roberts, J. M., Katsoulis, D. E., & Arnold, F. H. (2021). Biocatalytic transformations of silicon—The other group 14 element. *ACS Central Science*, 7(6), 944-953.

Sarai, N. S., Fulton, T. J., O'Meara, R. L., Johnston, K. E., Brinkmann-Chen, S., Maar, R. R., ... & Arnold, F. H. (2024). Directed evolution of enzymatic silicon-carbon bond cleavage in siloxanes. *Science*, 383(6681), 438-443.

Shanmugaiah, V., Gauba, A., Hari, S. K., Prasad, R., Ramamoorthy, V., & Sharma, M. P. (2023). Effect of silicon micronutrient on plant's cellular signaling cascades in stimulating plant growth by mitigating the environmental stressors. *Plant Growth Regulation*, 100(2), 391-408.

Shapiro, B. (2017). Pathways to de-extinction: how close can we get to resurrection of an extinct species?. *Functional Ecology*, 31(5), 996-1002.

Venter, J. C., Glass, J. I., Hutchison, C. A., & Vashee, S. (2022). Synthetic chromosomes, genomes, viruses, and cells. *Cell*, 185(15), 2708-2724.

Wang, S., Liu, X., & Zhou, P. (2022). The road for 2D semiconductors in the silicon age. *Advanced Materials*, 34(48), 2106886.

*Web of Science | Clarivate*. (n.d.). Academia and Government. https://clarivate.com/academia-government/scientific-and-academic-research/research-discovery-and-referencing/web-of-science/

Weng, M. M., Zaikova, E., Millan, M., Williams, A. J., McAdam, A. C., Knudson, C. A., ... & Johnson, S. S. (2022). Life underground: Investigating microbial communities and their biomarkers in Mars-analog lava tubes at Craters of the Moon National Monument and Preserve. *Journal of Geophysical Research: Planets*, 127(11), e2022JE007268.

White, P. J., & Brown, P. (2010). Plant nutrition for sustainable development and global health. *Annals of botany*, 105(7), 1073-1080.

Xu, Y., & Zhu, T. F. (2022). Mirror-image T7 transcription of chirally inverted ribosomal and functional RNAs. *Science*, 378(6618), 405-412.

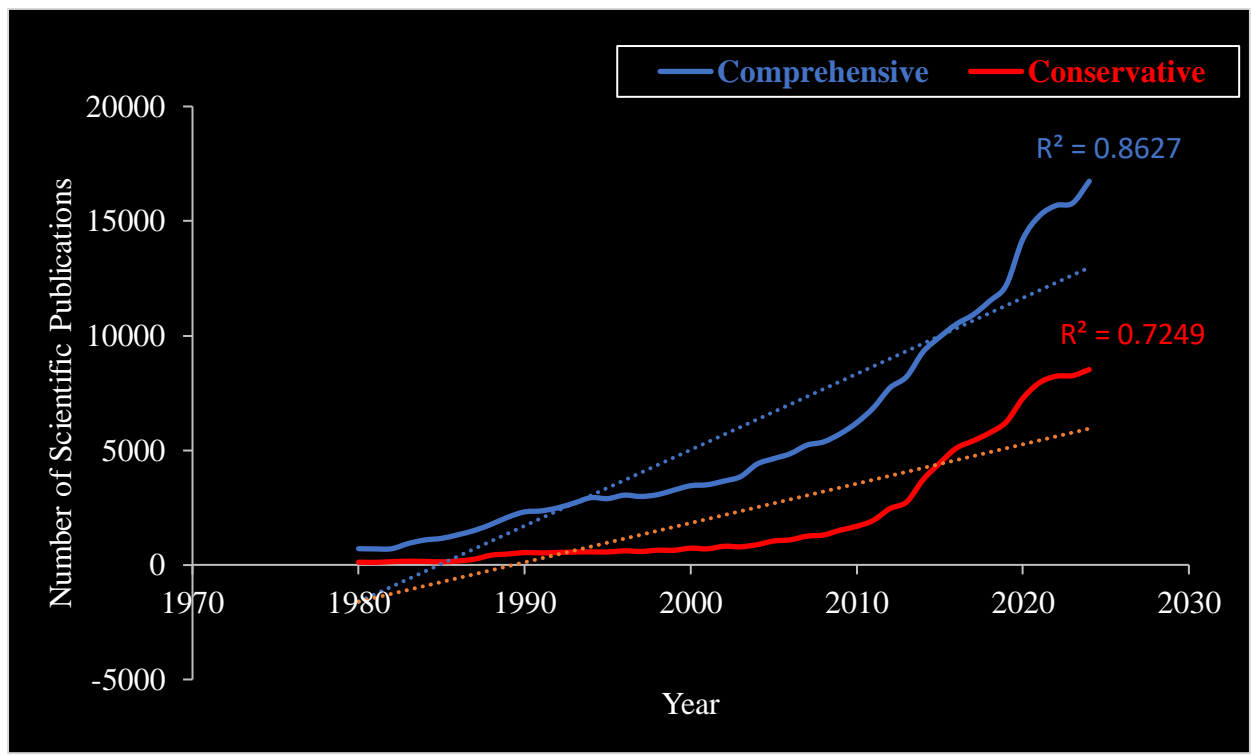

Figure 1

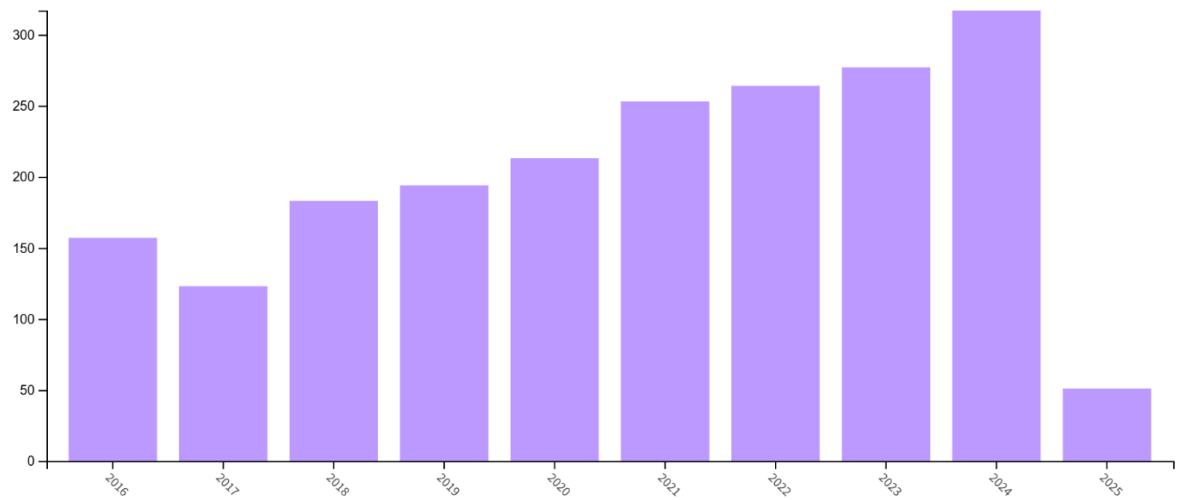

Figure 2 a

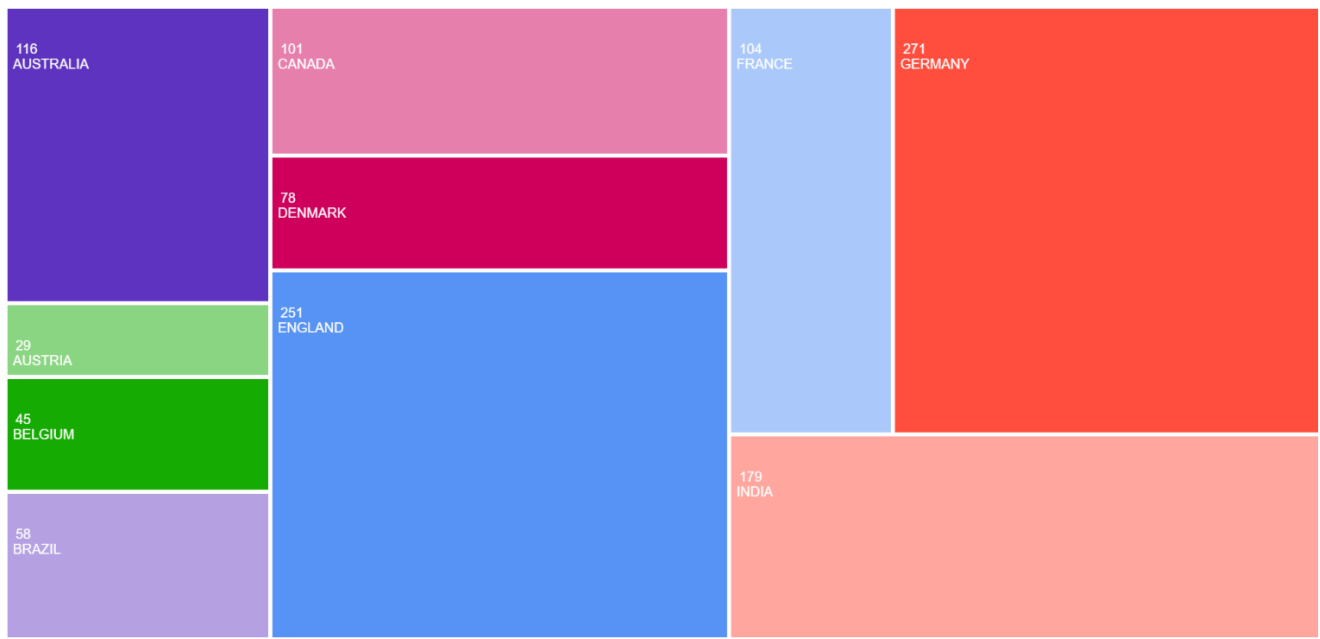

Figure 2 b

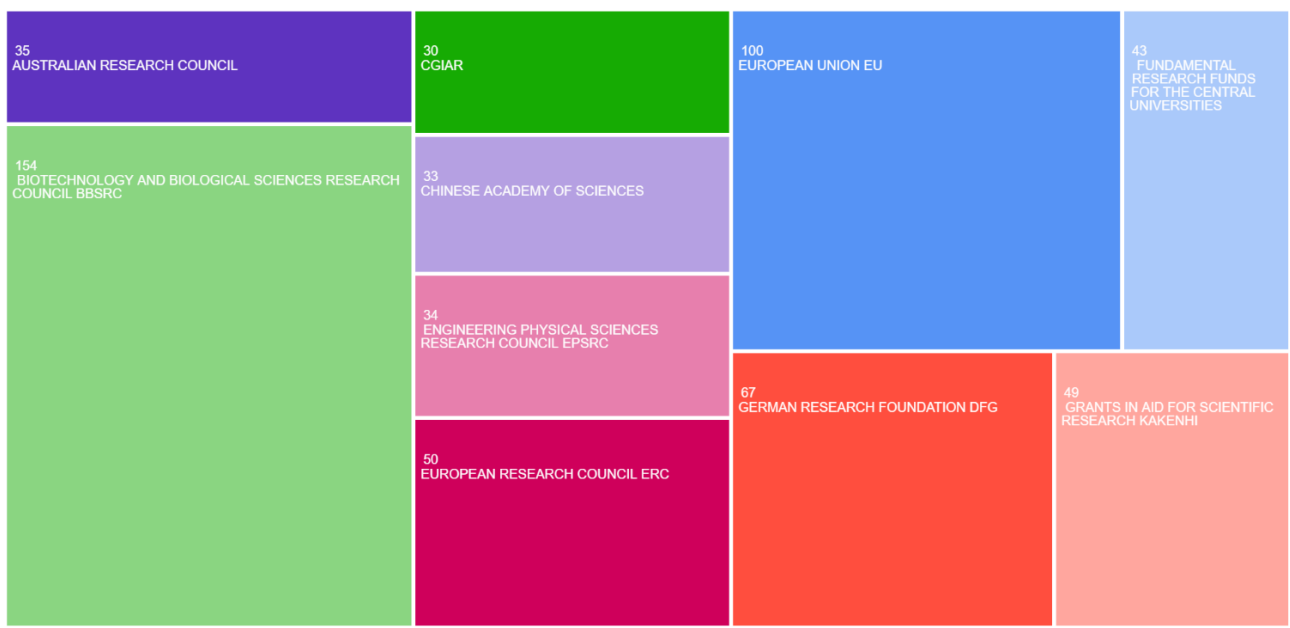

Figure 2 c

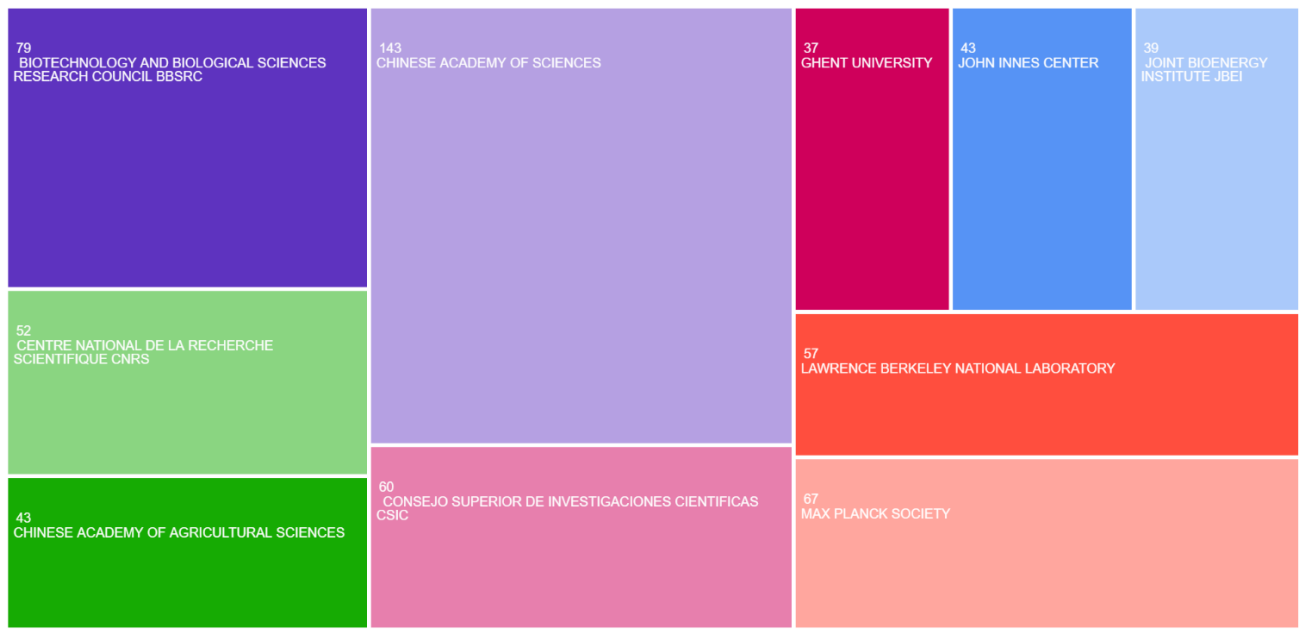

Figure 2 d.

| Article Type | Number of Publications |
|---|---|
| Research Support,Non-U.S. Gov't | 3218 |
| Review & Systematic Review | 1770 |
| Research Support, U.S. Gov't | 797 |
| Letter, Editorial & Comment | 142 |
| Comparative Study | 61 |
| Preprint | 20 |
| News & News Paper Article | 7 |
| Dataset, Technical Report | 7 |

Table 1

**Figure Legend**

**Figure 1**: Publication trends of silicon research and in relation to synthetic biology. Comprehensive search indicates a broader search category where search query includes silicon, artificial life, artificial cell, synthetic life, synthetic cell and synthetic biology. Conservative search indicates a more restrictive search category with terms strictly related to biological science, here the search query included silicon along with artificial life, synthetic life and synthetic biology.

**Figure 2 a:** Number of publications recorded in Web of Science from 2016 to March, 10th, 2025. The bar graph indicates a steadily increasing interest in the subject over a 10-years period.

**Figure 2 b:** A tree map representing the number of publications with respect to countries or region. Germany followed by England and India are top three countries with respect to publication numbers in plant synthetic biology.

**Figure 2 c:** A tree map indicating the major funding agencies that supports research in plant synthetic biology. Publication numbers indicate BBSRC or Biotechnology and Biological Sciences Research Council as the leading funding organization.

**Figure 2 d:** A tree map indicating the research organizations that generates most peer-reviewed articles on plant synthetic biology. Chinese Academy of Sciences followed by BBSRC and Max Planck Society are the three major organizations that invests on research related to plant synthetic biology.

**Table 1**: Plant synthetic biology publication trend from 2010 to 2024 was retrieved from PubMed. The search term "Plant Synthetic Biology" was used in the "All Fields" category. The different categories displayed in the table were curated manually.